\documentstyle[sprocl,epsf,axodraw]{article}

\def\Journal#1#2#3#4{{#1} {\bf #2}, #3 (#4)}
\def\IJMA{{\em Int. J. Mod. Phys.} A}
\def\RMP{\em Rev. Mod. Phys.}
\def\UPN{\em Usp. Fiz. Nauk}
\def\NPB{{\em Nucl. Phys.} B}
\def\PLB{{\em Phys. Lett.}  B}
\def\PRL{\em Phys. Rev. Lett.}
\def\PRD{{\em Phys. Rev.} D}

\def\be{\begin{equation}}
\def\ee{\end{equation}}
\def\ba{\begin{eqnarray}}
\def\ea{\end{eqnarray}}
\def\slash#1{#1\!\!\!/\!\,\,} 
\newcommand{\la}[1]{\label{#1}}
\newcommand{\nr}[1]{(\ref{#1})}
\newcommand{\lsi}{\raise0.3ex\hbox{$<$\kern-0.75em\raise-1.1ex\hbox{$\sim$}}}
\newcommand{\gsi}{\raise0.3ex\hbox{$>$\kern-0.75em\raise-1.1ex\hbox{$\sim$}}}
\newcommand{\lsim}{\mathop{\lsi}}
\newcommand{\gsim}{\mathop{\gsi}}
\newcommand{\nn}{\nonumber \\}

\newcommand{\fr}[2]{{\frac{#1}{#2}}}
\newcommand{\msbar}{\overline{\mbox{\rm MS}}}

\begin{document}
\vspace*{-2cm}
\begin{flushright}
HD-THEP-97-31\\
hep-ph/9707415\\
July 1997\\
\end{flushright}

\vspace*{0.5cm}

\title{3D EFFECTIVE THEORIES FOR THE\\ 
STANDARD MODEL AND EXTENSIONS}

\author{M. LAINE}

\address{Institut f\"ur Theoretische Physik,
Philosophenweg 16,\\ 
D-69120 Heidelberg, Germany}

\maketitle\abstracts{
The construction and analysis of 3d effective theories
for the description of the thermodynamics of the cosmological
electroweak phase transition are reviewed. For the Standard
Model and for the MSSM with stops somewhat heavier than the top, 
the effective theory is 3d SU(2)+Higgs, and the transition 
is strong enough for baryogenesis at no Higgs mass for the Standard Model,
and up to $m_H\lsim 80$ GeV for the MSSM. 
For lighter stops, the effective theory is SU(3)$\times$SU(2) + two scalars 
and the transition could be strong enough up to $m_H\sim 100$ GeV. 
However, a lattice study in 3d is needed to confirm the last bound.}

\section{Introduction}

The cosmological electroweak phase transition offers the prospect
of connecting perturbative physics measurable at accelerator experiments, 
such as the Higgs mass, to non-perturbative astrophysical consequences, 
such as the matter-antimatter asymmetry and the galactic magnetic fields.
Progress in recent years (for a review, see~\cite{rs}) has made at least 
parts of this connection quantitative. The purpose of this talk is to 
review the part which is perhaps best under control: the equilibrium
thermodynamics of the electroweak phase transition. 

More specifically, the physics questions under consideration are:
given the zero temperature vacuum Lagrangian and the corresponding
parameters (the Higgs mass $m_H$, etc), 
\begin{itemize}
\item
is there a phase transition at some critical temperature $T_c$? 
If there is one, what is the order of that transition?
\item 
if the transition is of the first order, what
are the latent heat, the surface tension, and the
discontinuity in the expectation value of the Higgs field
characterizing the transition?
\item
what are the correlation lengths of different
excitations around $T_c$?
\end{itemize}

Of course, the answers to
these questions are just the first step
in the study of the physical consequences of the transition.
The equilibrium real time processes --- such as the sphaleron
rate and the plasmon properties --- not to mention 
the whole class of non-equilibrium processes --- 
such as bubble growth, 
rates of different processes in the bubble wall background, 
CP-violation, etc --- remain outside of the present 
discussion. It is certainly
only these phenomena which constitute the real physics 
possibly responsible, e.g., for baryon number or
magnetic field generation.
However, knowing the equilibrium thermodynamics is 
a necessary starting point: it gives the ``ground state''
around which non-equilibrium phenomena take place.

In fact, there is also one important quantitative constraint
given (to a large extent) by thermodynamical considerations. 
This follows from the fact that if a baryon asymmetry has been
generated at the electroweak phase transition, 
it must be preserved afterwards until present day. As non-thermodynamical
input into this constraint go the sphaleron rate 
in the broken phase~\cite{krs,ar,khleb}, and the expansion rate
of the Universe (which is however determined by the equation
of state). The thermodynamical constraint to be satisfied
can then be expressed as~\cite{krs,klrs}
\be
\frac{v(T_c)}{T_c}\gsim 1.0\ldots 1.5, \la{voT}
\ee
where $v(T_c)$ is the expectation value of the Higgs field
in the broken phase in, say, the Landau gauge. 

Unfortunately, even solving for the equilibrium thermodynamics
of the electroweak phase transition
is quite a difficult task,
even though the theory is weakly coupled. This is due to the 
infrared (IR) problem at finite temperature~\cite{linde,gpy}.
Indeed, the perturbative expansion parameters of a finite
temperature system differ from those of the same system 
at zero temperature. If the vacuum theory has the coupling $g^2$, 
then at finite temperature some loops are, instead, 
proportional to 
\be
g^2 n_b(E) = \frac{g^2}{e^{E/T}-1}
\stackrel{E\ll T}{\sim}\frac{g^2T}{E}.
\ee
Here $n_b(E)$ is the Bose distribution function and $E$ is
the typical energy or mass scale of the particles contributing.
For example, in the effective potential $V(\phi)$ 
for the Higgs field $\phi$ appear terms of the type 
\be
\begin{array}{ccccc}
\frac{T}{4\pi}g^3\phi^3 & \frac{T^2}{(4\pi)^2}g^4\phi^2 &
\frac{T^3}{(4\pi)^3}g^5\phi & \frac{T^4}{(4\pi)^4} g^6 & \ldots \\
& & & & \\
({\rm\scriptstyle 1-loop}) & 
({\rm\scriptstyle 2-loop}) & 
({\rm\scriptstyle 3-loop}) & 
({\rm\scriptstyle 4-loop}) & 
\end{array}
\ee
For the Standard Model, 
the numerical coefficients in the 1-loop and 2-loop~\cite{ae} terms 
have been computed explicitly, and that of the 
3-loop term has been determined on the lattice~\cite{klrs}.
It follows that the expansion parameter is 
\be
\epsilon \sim \frac{g^2 T}{\pi m_W},
\ee
where $m_W\sim \fr12 {g \phi}$.
In perturbation theory the transition is between 
$\phi=0$ and $\phi\sim gT$; hence $\epsilon$ is not
small and 
the perturbative description might, a priori, be quite unreliable.
This estimate may get modified in extensions of
the Standard Model~\cite{cqw}, but especially
the symmetric phase always remains non-perturbative. 

It is hence clear that to solve for the 
equilibrium thermodynamics of the electroweak phase
transition, one needs a non-perturbative method. 
In principle the most straightforward is to put
the 4d theory on the lattice and to make 4d lattice
simulations~\cite{leip1}$^{-\,}$\cite{hein}.
Unfortunately, this turns out to be numerically quite demanding.
The purpose of this talk is to review another method:
dimensional reduction into 3d and lattice simulations 
in that theory. The discussion here is qualitative, and
more details can be found in the original papers~\cite{generic,klrs,su2u1} 
and reviews~\cite{rs,ms}. 

\section{Dimensional reduction}
\la{DR}

The basic idea of dimensional 
reduction~\cite{g}$^{-\,}$\cite{br}
is factorization: the difficult complete problem can be divided into
two easier parts, the first of which can be solved perturbatively
with good accuracy, and the second of which can be solved
on the lattice with reasonable computer resources. The 
perturbative step involves the integration out of momentum 
scales $p\gsim \pi T, gT$, and is free of IR-problems: 
the expansion parameters are just ${g^2}/{(4\pi)^2},
{g}/{(4\pi)}$. The numerical study is needed to 
investigate the dynamics of the 
soft bosonic modes $p\lsim g^2T$ with IR-problems.

To be somewhat more concrete, consider finite temperature
field theory in Matsubara formalism and with cutoff regularization. 
Then dimensional reduction for the Standard Model consists
of integrating out the following 
fields ($D$ and $S$ denote free propagators):
\be
1^{\rm st}\, {\rm step} 
\left\{
{\rm
\begin{minipage}{7.8cm}
\begin{itemize}
\item 
$n\neq0$ bosons, $D=\frac{1}{{\bf  p}^2+(\pi T)^2(2 n)^2+m^2}$
\item
all fermions, $S=\frac{i\slash{p}}{{\bf  p}^2+(\pi T)^2(2 n+1)^2}$
\item
modes with ${\bf  p}\gsim \pi T$ for $n=0$ bosons
\end{itemize}
\end{minipage}
}\right. \la{1st}
\ee
\be
2^{\rm nd}\, {\rm step} 
\left\{
{\rm
\begin{minipage}{7.8cm}
\begin{itemize}
\item 
$A_0$-field, $D=\frac{1}{{\bf  p}^2+m_D^2}, 
{\scriptstyle m_D^2\sim g^2T^2}$
\item
modes with ${\bf  p}\gsim m_D$ for other $n=0$ bosons
\end{itemize}
\end{minipage}
}\right. \la{2nd}
\ee
The $A_0$ field gets the mass $m_D$ radiatively 
by the integrations in the $1^{\rm st}$ step.
It should be stressed that all modes integrated
out in these steps are massive and thus there are
no IR-problems. 

As a result of these steps, one is left with 
a purely bosonic local effective 3d theory
for the $n=0$ Matsubara modes. The steps above 
were outlined in the language of a momentum cutoff, but the 
same steps can also be formulated in the $\msbar$ scheme. Then 
one does the reduction by a matching procedure~\cite{generic},
and the remaining effective theory is a 3d continuum field theory
without any restriction on the loop momenta.

The more formal statement for the reduction process
is as follows~\cite{generic}. Consider the standard electroweak
theory around the phase transition temperature $T\sim T_c$. Then
all static Green's functions for the ``light'' bosonic
(Higgs and gauge) fields can be obtained with relative 
error ${\delta G}/{G}\sim{\cal O}(g^3)$ from 
\be
{\cal L}_{\rm 3d} = 
{1\over4} G_{ij}^aG_{ij}^a+{1\over4} F_{ij}F_{ij}+
(D_i\phi)^\dagger D_i\phi+m_3^2\phi^\dagger\phi+
\lambda_3(\phi^\dagger\phi)^2, \la{L3d}
\ee
when $m_3^2,\lambda_3$ 
and the SU(2) and U(1) gauge couplings
$g_3^2,g_3'^2$ are suitably fixed.
Fixing can be done perturbatively in the $\msbar$ scheme.
The set of rules required for the fixing
at 2-loop level has been 
presented in~\cite{generic}.

The errors in ${\delta G}/{G}\sim{\cal O}(g^3)$
can basically arise from two sources. First, there are 
higher order corrections in the effective parameters, e.g., 
\be
m_3^2 \sim -\frac{m_H^2}{2}
\left[1+\biggl(\frac{g}{4\pi}\biggr)^2+\ldots\right]
+g^2T^2\left[1+\frac{g}{4\pi}+
\biggl(\frac{g}{4\pi}\biggr)^2+\ldots\right]. \la{paramerror}
\ee
Second, there are higher order operators, e.g., of the type  
\be
\left(\frac{m_{\rm small}^2}{m_{\rm large}^2}
\right)^{3+n} \sim
\biggl(\frac{g}{4\pi}\biggr)^m(\phi^\dagger\phi)^3,\ldots,
\la{opererror}
\ee
where $m_{\rm small}^2$ corresponds to a mass scale of a field
kept in the effective theory, $m_{\rm large}^2$ corresponds to 
a field integrated out, and $n\ge 0, m\ge 3$.

It is naturally one of the most essential points in 
the 3d approach to estimate what the parametric error
${\delta G}/{G}\sim{\cal O}(g^3)$
means numerically. For the physical Standard Model, 
the analytic estimates made~\cite{generic,jkp} show
that numerically the error should be on the $\lsim 5$\% level 
for the physical Higgs masses 
30 GeV $\lsim m_H \lsim$ 200 GeV. This
number certainly depends on the values 
of the parameters in the theory, and thus the errors
may be different in extensions. 

In principle, it would be nice to check the accuracy of
the reduction steps directly by a comparison with 4d simulations.
Unfortunately, the errors there are not at
a sufficiently small level, yet. In 
the comparisons made so far~\cite{ml2} (for a gauge coupling
larger than the physical value) the results 
were compatible within error bars. 

A key point is now that the relative 
error 5\% can indeed be reached in 3d simulations, 
with computer resources available at present. Thus the 
construction of an effective 3d theory  and lattice simulations
in that theory offer a way of obtaining reliable results for
the thermodynamical observables of
the physical electroweak phase transition.

Let us finally list some reasons why the 3d approach 
is a particularly economical way of obtaining reliable results,
in its realm of validity:

\begin{itemize}
\item
universality: 
the same 3d theory describes many different 4d theories, 
and thus their IR-problems can be solved once and for all.
In particular, the baryon number bound in eq.~\nr{voT}
can be converted to a property of the 3d theory 
(see Sec.~\ref{SM}), and thus the non-perturbative 
properties of all the theories which
reduce to eq.~\nr{L3d} can be handled simultaneously. 

\item
the effective 3d theory
automatically implements the resum\-ma\-tions \linebreak
needed at finite temperature.

\item 
in the 3d approach, 
zero-temperature renormalization 
is made in the perturbative reduction step, 
and thus connection to 4d physics
(the top quark mass, the muon lifetime/$\msbar$ gauge coupling, etc.) 
can be made very easily, unlike in direct 4d simulations.

\item
the 3d theory is super-renormalizable, 
allowing a relatively easy continuum extrapolation
in the lattice simulations.

\item
there are only a few length scales in the 3d theory, 
so that the lattice sizes need not be prohibitively large.
Indeed, to describe continuum physics, the lattice spacing
$a$ has to be smaller than the smallest correlation length
in the system
and the lattice size $Na$ has to be larger than the largest.
Since the scales in parentheses below do not appear any more
in the effective theory, sufficiently large values of $N$ can be 
chosen: 
\be
a \underbrace{
\left[ \ll \frac{1}{\pi T} \ll \frac{1}{\sqrt{2}gT}\right]}_{
{\rm do\,\, not\,\, appear\,\, in\,\, eq.\nr{L3d}}}
\ll \frac{1}{g^2T} \ll Na.
\ee

\end{itemize}

\section{The Standard Model}
\la{SM}

The program of dimensional reduction
outlined in the previous Section
has been applied
to the electroweak sector of the Standard Model 
in~\cite{generic,jkp}, and the required 
3d lattice simulations have been 
carried out in~\cite{kars}$^{-\,}$\cite{mt}.
The implications of the lattice results for the 
sphaleron bound of eq.~\nr{voT} have been considered in~\cite{klrs}.
Let us review the main results here. 

\subsection{Reduction to 3d SU(2)$\times$U(1)+Higgs}
\la{SMDR}

The first task is the construction of the effective theory
in eq.~\nr{L3d}. The main steps are, 
following eqs.~\nr{1st}, \nr{2nd}: 

1. Renormalization of the 
vacuum Lagrangian in the $\msbar$ scheme in terms of the
physical parameters of the theory. For the Standard Model, 
these are essentially $m_H,m_Z$, $m_{\rm top}$, $\tau_\mu$,
$\alpha_{\rm EM}$, $\alpha_{\rm QCD}$, where $\tau_\mu$ is 
the muon lifetime. Symbolically, at 1-loop level this step
corresponds, e.g., to 

\be
\begin{minipage}{1.6cm}
\begin{picture}(40,20)(0,0)
\CArc(20,10)(10,0,360)
\DashLine(0,0)(40,0){5}
\Text(20,-5)[]{$\scriptstyle g^2$}
\Text(30,20)[lb]{$\scriptstyle T=0$}
\end{picture}
\end{minipage}
\hspace*{1.0cm} \Rightarrow
m^2(\mu)=-\frac{m_H^2}{2}+\Pi(-m_H^2,\mu),
\hspace*{2.1cm}
\ee

\vspace{0.2cm}

\noindent
where the dashed line represents the Higgs field
and $m_H$ is the pole Higgs mass.

2. Dimensional reduction 
in the $\msbar$ scheme by matching
Green's functions in the 4d theory and in the 3d theory, 
corresponding to eq.~\nr{1st}.
Symbolically, the loops are at 1-loop level of the type

\be
\begin{minipage}{1.6cm}
\begin{picture}(40,20)(0,0)
\CArc(20,10)(10,0,360)
\DashLine(0,0)(40,0){5}
\Text(0,-5)[]{$\scriptstyle n=0$}
\Text(40,-5)[]{$\scriptstyle n=0$}
\Text(20,-5)[]{$\scriptstyle g^2$}
\Text(30,20)[lb]{$\scriptstyle n\neq 0$}
\end{picture}
\end{minipage}
\hspace*{1.0cm} \Rightarrow
m_3^2=g^2(\mu)T^2 +m^2(\mu)
\left[
1+\frac{g^2}{(4\pi)^2}\ln\frac{\mu}{2\pi T}
\right]. \la{drm32}
\ee

\vspace{0.2cm}

\noindent
This computation should be done at the 2-loop level
to fix the scale appearing in $g^2(\mu)T^2$.

3. Integration over the 3d heavy scales 
in the $\msbar$ scheme by matching, 
corresponding to eq.~\nr{2nd}. 
For the Standard Model, the heavy scale
is given by the zero component of the gauge field $A_0$,
with the Debye mass $m_D\sim gT$. Symbolically, 

\be
\begin{minipage}{1.6cm}
\begin{picture}(40,20)(0,0)
\CArc(20,10)(10,0,360)
\DashLine(0,0)(40,0){5}
\Text(0,-5)[]{$\scriptstyle n=0$}
\Text(40,-5)[]{$\scriptstyle n=0$}
\Text(20,-5)[]{$\scriptstyle g_3^2$}
\Text(30,20)[lb]{$\scriptstyle A_0$}
\end{picture}
\end{minipage}
\hspace*{1.0cm} \Rightarrow
m_3^{2({\rm new})}=m_3^{2({\rm old})}-\frac{g_3^2}{16\pi} m_D.
\hspace*{2.3cm}
\ee

\vspace{0.2cm}

\noindent
As a result of this step, one gets the effective theory
in eq.~\nr{L3d}.

4. Finally, one should change scheme from $\msbar$ to lattice
regularization. In the theory of eq.~\nr{L3d}, this can be 
done exactly (in the continuum limit) with a 2-loop 
computation~\cite{fkrs2,ml,lr}. One can also remove
most of the $O(a)$-effects~\cite{gdm}, making the  
extrapolation to the continuum limit less costly. After
these analytical steps, one can simulate the theory 
on the lattice. 

\subsection{Lattice simulations of 3d SU(2)$\times$U(1)+Higgs}
\la{xydef}

When doing the simulations, 
one can forget for a moment about the expressions
of the 3d parameters in terms of the physical 4d parameters, 
described in Sec.~\ref{SMDR}.
Indeed, the effective
theory in eq.~\nr{L3d} has four parameters
($g_3^2$, $g_3'^2$, $m_3^2$, $\lambda_3$), and as all of
them are dimensionful, one can choose one (say, $g_3^2$)
to fix the scale. Then the dynamics of the 3d theory 
only depends on three dimensionless parameters. However, 
the properties of the phase transition depend essentially
only on one parameter. This is because
\ba
\bullet\quad 
z & \equiv & \frac{g_3'^2}{g_3^2}\sim 0.3 \,\,\mbox{\rm is fixed by the
physical value of the Weinberg angle},  \nn
\bullet\quad 
y & \equiv  & \frac{m_3^2(g_3^2)}{g_3^4}
\sim c_1\frac{T-T_c^{\rm pert}}{T_c^{\rm pert}} \,\,\mbox{\rm 
is tuned to find the transition}, \nn
\bullet\quad 
x & \equiv & \frac{\lambda_3}{g_3^2}\sim 
\fr18\frac{m_H^2}{m_W^2}+c_2\frac{m_{\rm top}^4}{m_W^4} \,\,
\begin{minipage}[t]{6cm}
is the parameter left, 
determining the properties of the transition. 
\end{minipage}
\nonumber
\ea

The simulations are based on measuring gauge invariant composite
operator expectation values and distributions, e.g., 
$\langle \phi^\dagger\phi\rangle$, $P(\phi^\dagger\phi)$, etc.
From these one can determine various
physical quantities. For instance, 
the latent heat is related to the two-peak structure seen
in a first order transition, 
$L\propto \Delta\langle\phi^\dagger\phi\rangle$, 
and the surface tension is related to how the height of the 
plateau between the two peaks scales with the cross-sectional
area of the lattice. In Fig.~\ref{sims}, a set 
of typical results from lattice simulations are shown.

\begin{figure}[p]

\vspace{-1.0cm}

\centerline{\hspace{-3.3mm}
\epsfxsize=8cm\epsfbox{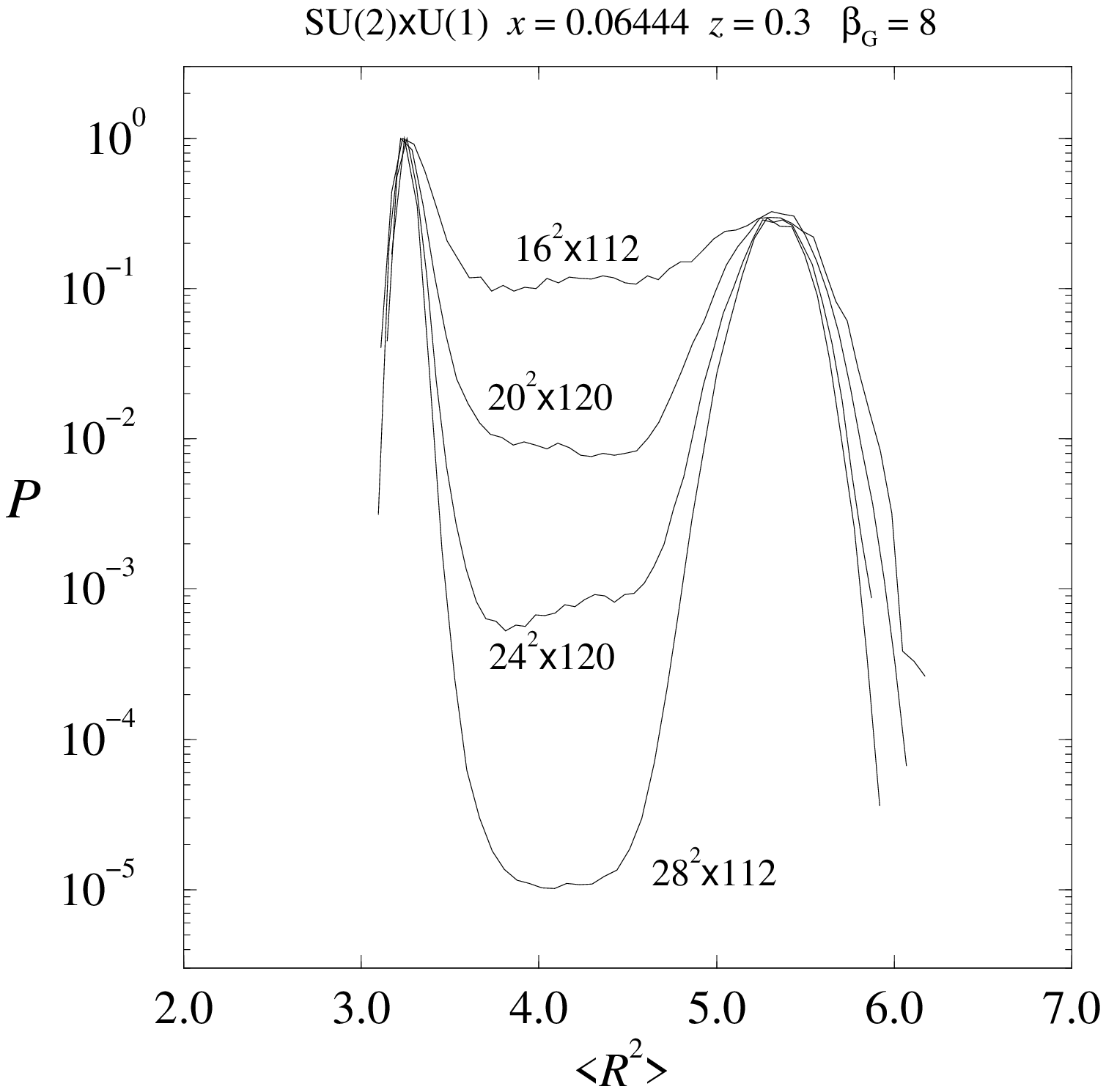}
\hspace{-1cm}
\epsfxsize=8cm\epsfbox{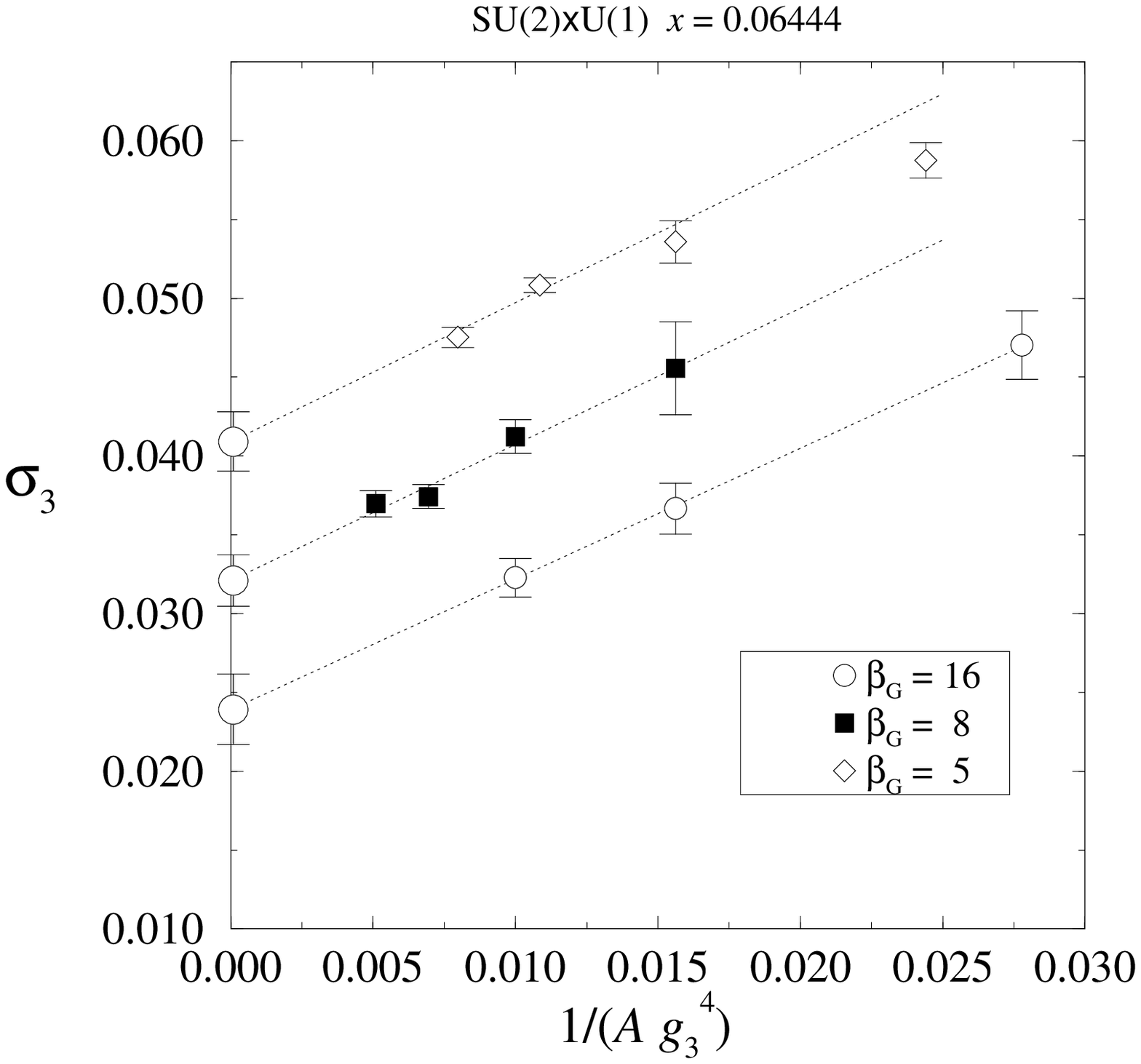}}

\vspace{-3.5cm}

\centerline{\hspace{-3.3mm}
\epsfxsize=8cm\epsfbox{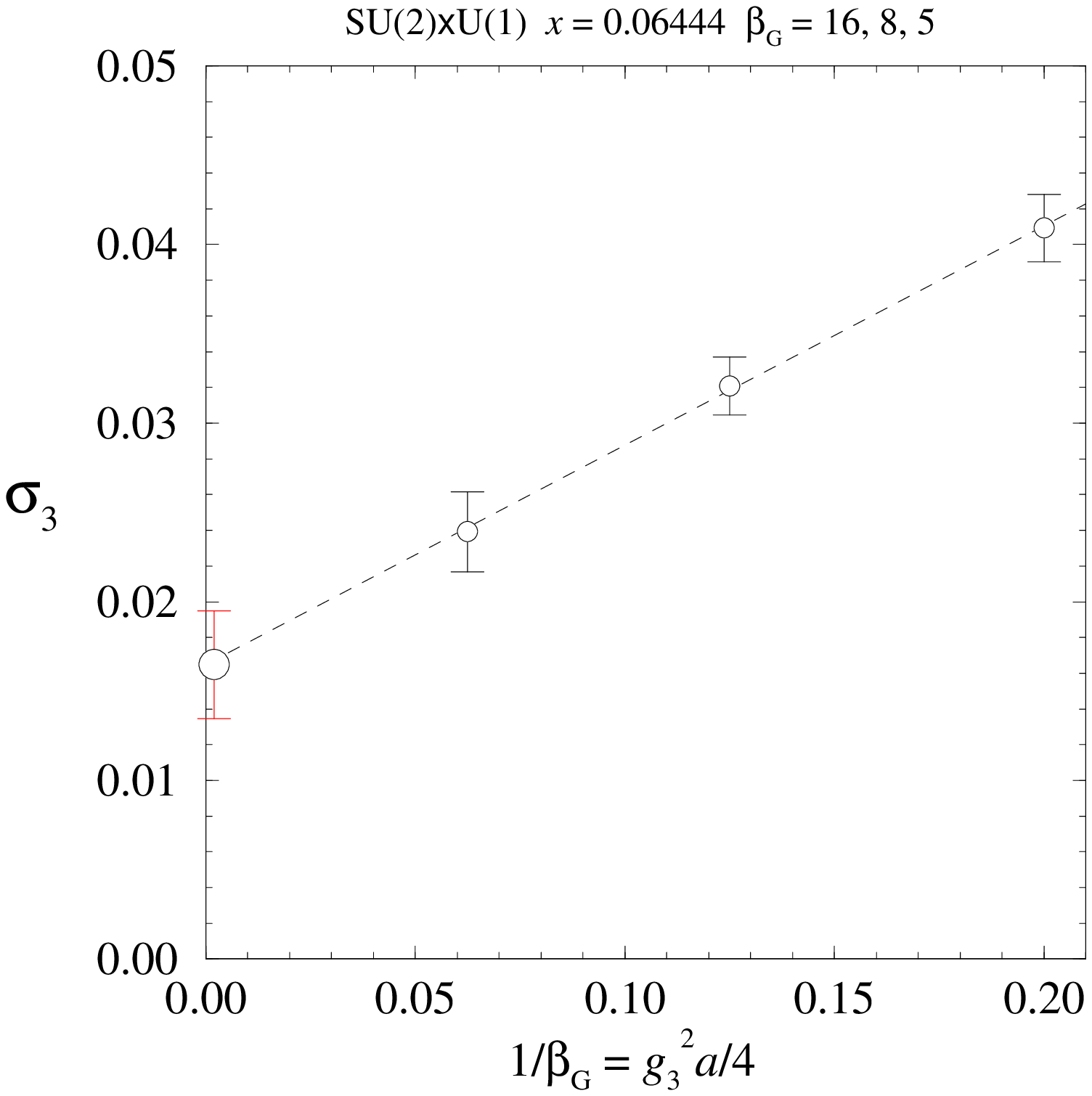}
\hspace{-1cm}
\epsfxsize=8cm\epsfbox{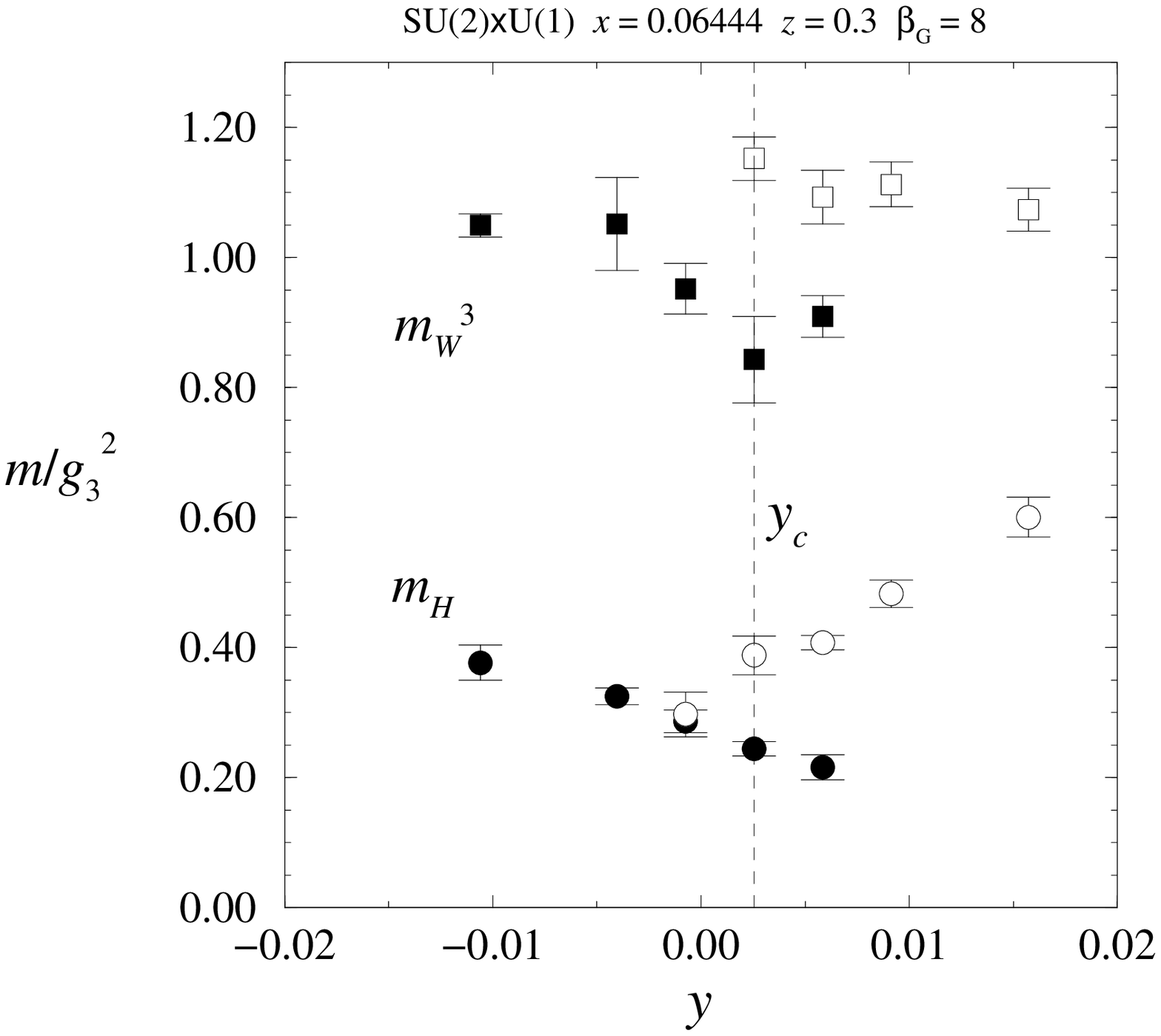}}

\vspace*{-11.3cm}

\centerline{\hspace{1cm} (a) \hspace{6.5cm} (b)}

\vspace*{7cm}

\centerline{\hspace{1cm} (c) \hspace{6.5cm} (d)}

\caption[a]{Lattice results for the 3d SU(2)$\times$U(1)+Higgs model
(from~\cite{su2u1}).
(a) The two-peak structure indicating a first order transition.
The distance between the two peaks determines the latent heat and
the  height of the plateau between the peaks the surface tension. 
(b) The infinite volume extrapolation for the surface tension. 
(c) The continuum extrapolation for the surface tension.
(d) The correlator masses around the transition point.}
\la{sims}
\end{figure}

The results of the lattice simulations for the 
phase diagram are summarized in Fig.~\ref{xyc}
(this figure is actually for the SU(2)+Higgs model without
the U(1) factor; the difference is small~\cite{su2u1}).
The main feature is that at small $x$, there is 
a first order phase transition which gets weaker
as $x$ increases, as predicted by perturbation theory.
However, unlike in perturbation theory,  the non-perturbative
transition line ends at $x\sim 0.11$. To the right of the endpoint, 
there is no first or second order transition. 

\begin{figure}[tbh]

\vspace*{-1.5cm}
 
\epsfxsize=10cm
\centerline{\epsffile{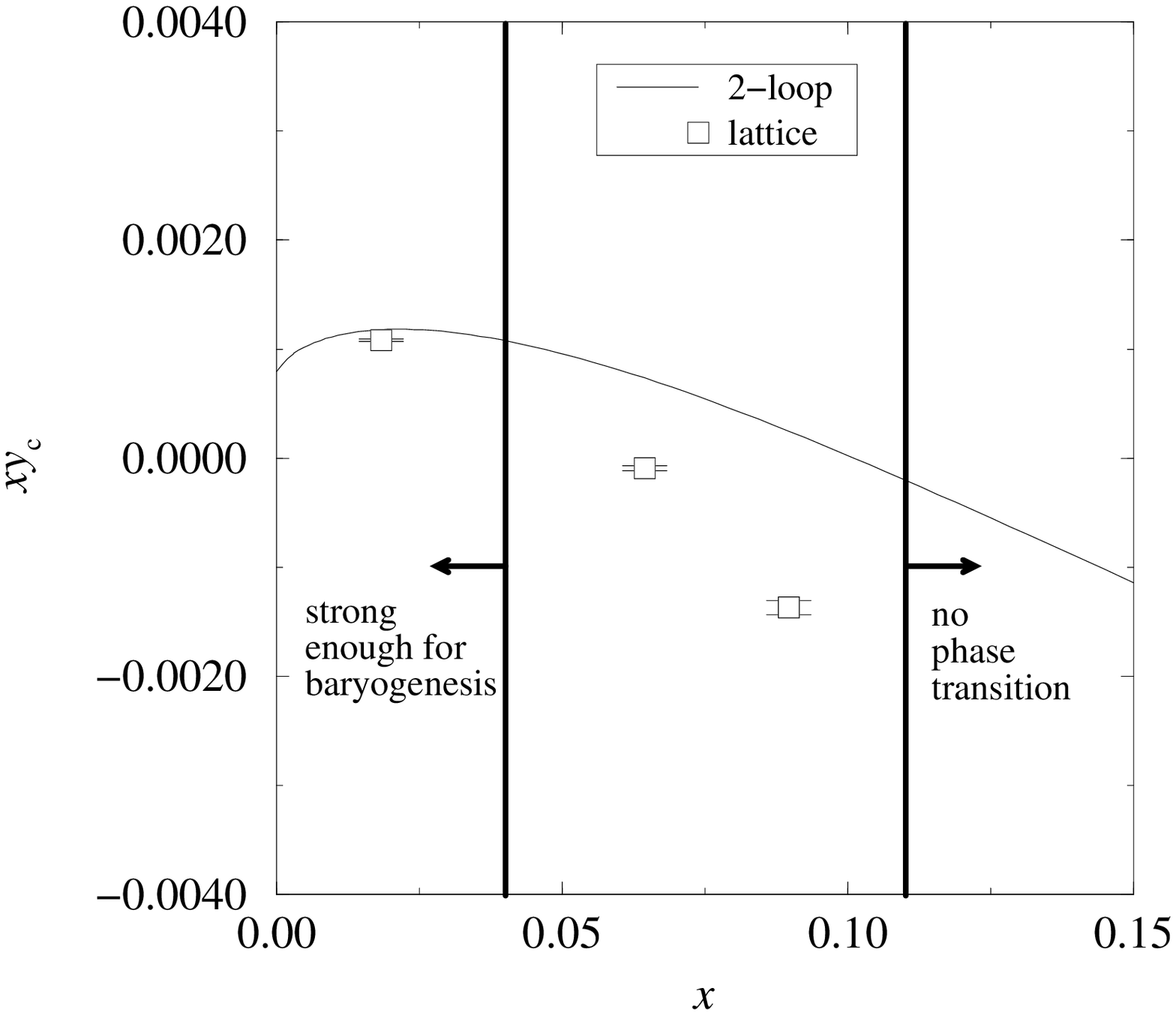}}
 
\vspace*{-5.0cm}

\caption[a]{The phase diagram of the 3d SU(2)+Higgs model~\cite{klrs}.
The parameters $x,y$ are defined in Sec.~\ref{xydef}. 
``Strong enough'' refers to the constraint in eq.~\nr{x}.}
\la{xyc}
\end{figure}

In Fig.~\ref{xyc}, there is also information
about the strength of the transition in the first order
regime. Indeed, whether eq.~\nr{voT} is satisfied or
not, depends on the non-perturbative dynamics of 
the effective theory. It has been shown in~\cite{klrs}
that eq.~\nr{voT} is satisfied provided that 
\be
x\lsim 0.03\ldots0.04. \la{x}
\ee 
This is a useful result because of its universal
character: if there is some other 4d theory
than the Standard Model, having the same effective Lagrangian
with the same parameter values, then the IR dynamics 
will be the same! Thus it is sufficient to derive 
the effective theory perturbatively. The non-perturbative
IR-features have already been accounted for by 
replacing the constraint in 
eq.~\nr{voT} by eq.~\nr{x}.

Finally, one can plug in 
the values of the 4d physical parameters
into the expressions of the 3d parameters. 
The result is shown
in Fig.~\ref{msm}, where also the lattice
results from Fig.~\ref{xyc} are included.
It is seen that the electroweak phase transition
in the Standard Model is never strong enough
for baryogenesis. It is of first order, however, 
for $m_H \lsim 70-80$ GeV.

\begin{figure}[tbh]

\vspace*{-1.5cm}
 
\epsfxsize=10cm
\centerline{\epsffile{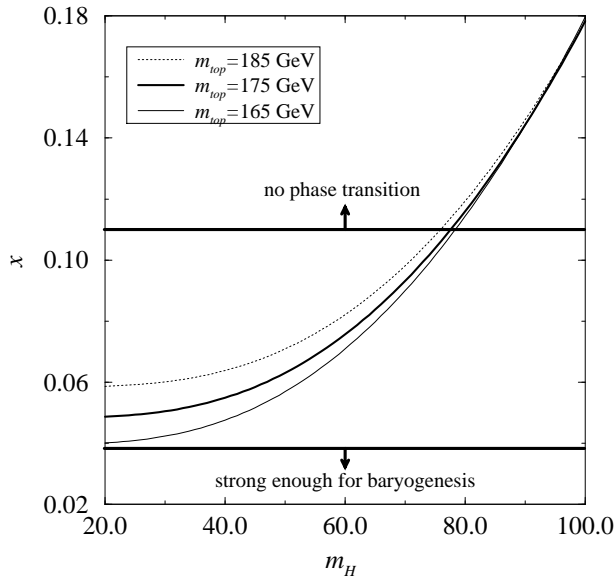}}
 
\vspace*{-5.0cm}

\caption[a]{The parameter $x$ determining the strength of the 
phase transition for the Standard Model, as a function of the 
pole Higgs mass $m_H$ and the pole top mass $m_{\rm top}$ (from~\cite{klrs}).}
\la{msm}
\end{figure}

\section{MSSM}

Let us then turn to the MSSM. The thermodynamics of the 
electroweak phase transition in the MSSM has recently been
considered perturbatively in~\cite{cqw}$^{-\,}$\cite{esp3}. 
Dimensional reduction has been applied to the MSSM
in~\cite{mssm}$^{-\,}$\cite{cl2} (dimensional 
reduction has recently been applied to a 
simple GUT-model, as well~\cite{su5}).

The main observation now is that for a part of the parameter
space of the MSSM, the effective theory is the same as in 
the Standard Model, i.e., given by eq.~\nr{L3d}. Then one 
immediately knows that the IR dynamics is the same, and in 
particular, that an IR-safe characterization of the strength of
the transition can be obtained by computing the parameter $x$.
On the other hand, one can also find cases where the IR-dynamics
of the transition is more complicated; then another effective
3d theory should be studied on the lattice. 

\subsection{Reduction to SU(2)$\times$U(1)+Higgs}

Let us first see how it could be that the effective theory
for the MSSM is just the one given by eq.~\nr{L3d}. At first sight, 
this may not be obvious since the field content of the 4d theory
differs quite a lot from the Standard Model. To be more specific, 
the minimal field content to be considered in the MSSM is that 
one has two Higgs doublets 
$H^1$, $H^2$; 
the SU(2) and SU(3) gauge fields
$A^a_\mu$, $C^A_\mu$;
the fermionic superpartners of the mentioned fields, i.e., 
higgsinos and gauginos; 
quarks of the $3^{\rm rd}$ generation 
$q_L,t_R,b_R$; 
the scalar superpartners of these, i.e.,
squarks of the $3^{\rm rd}$ generation
$Q, U, D$. 

To arrive at eq.~\nr{L3d}, one now does the following:
\begin{itemize}
\item 
renormalize the vacuum theory analogously to the
Standard Model case. Instead of $m_H$, the Higgs 
sector is usually parameterized in terms of 
$\tan\!\beta$ and the CP-odd Higgs mass $m_A$.
\item
go to finite temperature and 
integrate out all $n\neq0$ modes;
in particular, all the fermions are integrated out. 
This removes all the fermionic 
superpartners (gauginos and higgsinos)
from the effective theory.
\item
integrate out the zero components of the SU(2) and
SU(3) gauge fields, $A_0$, $C_0$, and the
heavy squark fields in 3d. 
In the generic case, all the squark fields are heavy, since
the electroweak transition takes place in the Higgs direction.
Thus only the gauge fields and two Higgs doublets are left.
\item
diagonalize the two Higgs doublet model
at the transition point where the mass
matrix has an eigenvalue close to zero.
\item
integrate out the heavy Higgs doublet. Since
the sum of the mass eigenvalues is 
$\sim m_1^2(T)+m_2^2(T)\sim m_A^2 + c T^2$
and since one of the eigenvalues is close to zero, 
the other Higgs field is heavy 
(especially if $m_A$ is not very small).
\end{itemize}
As a result, one indeed gets the effective theory
in eq.~\nr{L3d}.

However, in every new application, one has to be
careful with the accuracy of the reduction steps.
Let us therefore see what the expansion parameters
discussed in eqs.~\nr{paramerror},
\nr{opererror} look like for the MSSM.

1. For step 1 of eq.~\nr{1st}, 
the largest errors of the type in eq.~\nr{paramerror} correspond to
strong coupling constant and Yukawa corrections:
\be
(N_c^2-1)\frac{g_S^2}{(4\pi)^2}\ll 1, \hspace*{0.5cm}
N_c \frac{h_t^2}{(4\pi)^2}\ll 1.
\ee
These should be relatively well under control.
It should be noted that this kind of terms appear
more frequently in the MSSM than in the Standard Model.

2. For step 1 of eq.~\nr{1st}, eq.~\nr{opererror} 
corresponds basically to 
the high-tempe\-ra\-ture expansion for the fields that 
are kept in the effective theory. There are 
many more dimensionful parameters in the MSSM 
than in the Standard Model, so that one should require
\be
\frac{m_A^2}{(\pi T)^2},
\frac{m_Q^2}{(\pi T)^2},
\frac{m_U^2}{(\pi T)^2},
\frac{\tilde{A}_t^2}{(\pi T)^2} \ll 1, \ldots .
\ee
If these conditions are not satisfied for some field, 
then it has to be treated differently. 
Another part of eq.~\nr{opererror} is that the transition
should not be exceedingly strong. 

3. For step 2 of eq.~\nr{2nd}, 
the largest errors in eq.~\nr{paramerror}
correspond again to strong coupling constant and Yukawa corrections. 
However, in the 3d theory, the mass parameters also appear
for dimensional reasons:
\be
\frac{h_t^2 T}{\pi m_{U3}}, 
\frac{g_S^2 T}{\pi m_{U3}},
\frac{h_t^2 T}{\pi m_{Q3}}, 
\frac{g_S^2 T}{\pi m_{Q3}}\ll 1. \la{stop} 
\ee
There are also constraints for the mixing parameters, 
something like
\be
\frac{|\mu|}{m_{Q3}},
\frac{|{A}_t|}{m_{Q3}}\ll 1. \la{mix}
\ee
These constraints have been 
discussed in~\cite{cl1}
(see also Sec.~5 in~\cite{mssm}).

4. For step 2 of eq.~\nr{2nd}, eq.~\nr{opererror} is related to 
the strength of the phase transition and, provided
that eq.~\nr{stop} is satisfied, should be under control
unless the transition is exceedingly strong ($v/T\gsim 3$), 
see~\cite{bjls}.

Of these error estimates, the ones first
causing problems are eqs.~\nr{stop},
\nr{mix}. In particular, in the small stop 
scenario~\cite{cqw,dggw} 
the mass parameter $m_{U3}$ is small, 
$m_{U3}\sim 0$, and the first two
expansion parameters in eq.~\nr{stop} 
break down. Then the $U$-field cannot be integrated out, 
and one has to go to the treatment in Sec.~\ref{drsu3su2}.  
The second problem may be associated with 
eq.~\nr{mix} in the case that there is large mixing in 
the squark sector~\cite{cl1}.

On the other hand, there are certainly cases where 
the expansion parameters {\em are} small. One typical 
such case is shown in Fig.~\ref{mssm}. 
Here the physical right-handed stop mass is 
$m_{\widetilde{t_R}}^2\approx m_U^2+m_{\rm top}^2$, 
so that $m_U\approx 50$ GeV corresponds to 
$m_{\widetilde{t_R}}\approx 180$ GeV. It is seen
that one can indeed get a strong enough phase transition
if $m_A$ is large and $m_H\lsim 80$ GeV. 
On the other hand, the transition is getting
stronger as $m_U$ is becoming smaller, in which 
limit the reliability of the reduction is decreasing
due to the fact $m_{U3}$ in eq.~\nr{stop} is becoming
smaller. However, for $m_U=50$ GeV the expansion 
parameters are still relatively small. Hence the conclusion 
is that one can have a transition strong enough 
for baryogenesis for the experimentally allowed 
Higgs masses, unlike in the 
Standard Model~\cite{cqw}$^{-\,}$\cite{cl2}. 


Finally, though not indicated
in Fig.~\ref{mssm}, it is interesting to note that 
for these parameters one always gets a first order 
transition as one increases $\tan\!\beta$ (or $m_H$),
although the transition is getting weaker. The situation 
is thus quite different from the Standard Model case, 
where there is no transition at high enough Higgs masses.

\begin{figure}[htb]

\vspace*{-1.5cm}
 
\epsfxsize=10cm
\centerline{\epsffile{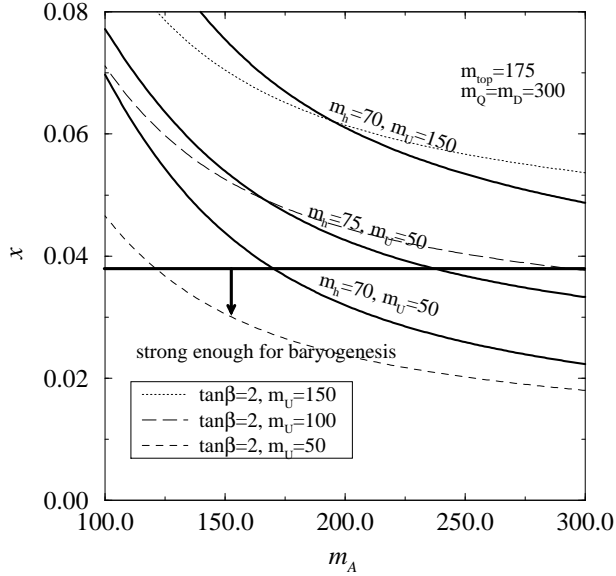}}
 
\vspace*{-5.0cm}

\caption[a]{The parameter $x$ in the MSSM, in terms of
the CP-odd Higgs mass $m_A$, the lightest CP-even Higgs 
mass $m_h$ and the stop mass parameter $m_U$ (from~\cite{mssm}). The 
physical right-handed stop mass is 
$m_{\widetilde{t_R}}^2\approx m_U^2+m_{\rm top}^2$, 
so that $m_U\approx 50$ GeV implies
$m_{\widetilde{t_R}}\approx 180$ GeV.}
\la{mssm}
\end{figure}

\subsection{Reduction to SU(3)$\times$SU(2)+ two scalars}
\la{drsu3su2}

Let us then consider the light stop scenario. 
In that case that the right-handed stop cannot be 
integrated out, since the expansion parameters in 
eq.~\nr{stop} would explode. Thus the stop 
field $U$ has to be kept in the
effective theory. The form of the effective 
3d theory can be written down, 
based on gauge invariance~\cite{mssm}: 
\ba
{\cal L}_{\rm 3d} & = &
\fr14 F^a_{ij}F^a_{ij}+\fr14 G^A_{ij}G^A_{ij}+ 
\gamma_3 H^\dagger H U^\dagger U \nn
& + & (D_i^w H)^\dagger(D_i^w H)+m_{H3}^2 H^\dagger H+
\lambda_{H3} (H^\dagger H)^2 \nn 
& + & (D_i^s U)^\dagger(D_i^s U)+m_{U3}^2U^\dagger U+
\lambda_{U3} (U^\dagger U)^2. \la{su3L3d}
\ea
Here $D_i^w$ and $D_i^s$ are the 
SU(2) and SU(3) covariant derivatives, and 
$F_{ij}^a$, $G_{ij}^A$ are the corresponding field strengths.
The U(1) group was for simplicity neglected. 
The form in eq.~\nr{su3L3d} assumes, e.g., 
that the squark  
mixing parameters are not exceedingly large.

The values of the parameters in eq.~\nr{su3L3d}
have been computed at 1-loop level
in a particular scenario in~\cite{bjls}, with
an estimate of the most important 2-loop effects included.
It should be stressed that the 2-loop effects 
are numerically quite important in some cases
and a complete
2-loop derivation might be useful if the MSSM
parameters turn out to be such that the effective
theory in eq.~\nr{su3L3d} is relevant. 
This is because the scale of the couplings 
in the thermal screening terms is fixed only at
the 2-loop level as discussed after eq.~\nr{drm32}, 
and for the squark mass parameters where
the strong gauge coupling appears,
this can have a significant numerical effect.

On the other hand, independent of the 
accuracy of the reduction steps, 
it would be quite important to analyze the theory
of eq.~\nr{su3L3d} on the lattice. 
This is because the IR-modes of eq.~\nr{su3L3d} are exactly 
those which are responsible for the strengthening 
effects seen in~\cite{cqw}$^{-\,}$\cite{esp2}. Unfortunately, no lattice
results are available at the moment.

In perturbation theory, 
the possible transitions in the theory of eq.~\nr{su3L3d} 
have been studied 
at the 2-loop level in~\cite{bjls}. 
Using the reduction formulas for the 3d parameters
in terms of 4d physics, one can see that 
the electroweak phase transition can indeed be quite
strong for small right-handed stop masses. Results 
for the phase diagram and for the constraint of eq.~\nr{voT}
are shown in Fig.~\ref{su3su2}.
It is seen that Higgs masses up to 100 GeV could be allowed
and that there might even be the
possibility of a two-stage transition. 

However, it should be noted that
there is a relatively large gauge
and scale dependence in the 2-loop 
results in this regime~\cite{bjls}.
Thus, even if the transition is strong (which in the 
Standard Model case turned out to imply that the IR-problems
are not very severe, after all), there is some uncertainty
in the perturbative results in the present case and a lattice
study is really needed. 

Finally, concerning the lattice study, let us note 
that there are many more parameters in eq.~\nr{su3L3d}
than in eq.~\nr{L3d}. In fact, if the strong gauge 
coupling $g_{S3}^2$ is chosen to fix the scale, then there
are six dimensionless parameters left. Thus one
has in a sense less universality than in eq.~\nr{L3d}, 
and one has to choose somewhat more specific 4d parameter
values for the lattice study.

\begin{figure}[tb]

\vspace*{-1.5cm}
 
\epsfxsize=10cm
\centerline{\epsffile{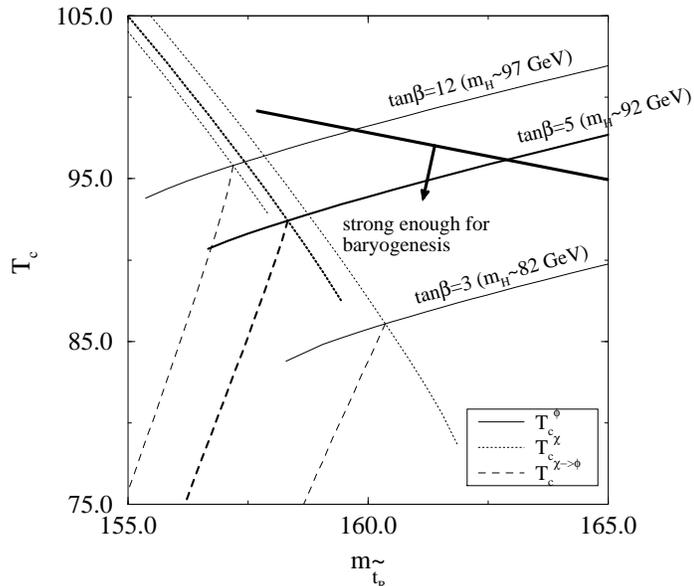}}
 
\vspace*{-5.0cm}

\caption[a]{The phase structure of the MSSM with a light
stop according to~\cite{bjls}. The solid lines indicate the 
conventional SU(2) transition, whereas the dotted lines
indicate a transition to the colour breaking stop direction.
The dashed lines are transitions from the stop direction
to the standard SU(2) broken minimum. Within the approximations 
made and for the parameters used, it thus appears possible,
in principle, to have a two-stage transition.}
\la{su3su2}
\end{figure}

\section{Conclusions and Outlook}

The method of dimensional reduction allows the construction 
of effective 3d theories for the electroweak sector of the
Standard Model, as well as for many extension thereof. This
construction is free of IR-problems and perturbative, and thus
at least in the cases where perturbation theory works at zero
temperature, one expects the reduction to be reliable. However, 
one should keep in mind that there are also other expansion 
parameters involved, especially related to 3d heavy scale
integrations. Thus it is important to estimate the accuracy
of the reduction steps each time a new theory is studied. 

In cases where the expansion parameters of the analytic reduction 
steps are getting larger, it would, in principle, be nice to have 4d
simulations for comparison. The realistic full 4d 
theory is far too complicated for this, 
but one might hope that in somewhat simplified models
(like the pure 4d SU(2)+Higgs theory for the Standard Model),
one could eventually make a meaningful comparison. 
Moreover, the accuracy of 3d heavy scale integrations
could be checked non-perturbatively
with 3d simulations in a more complete
effective theory, which is still less demanding than full 4d simulations.

Once a 3d theory has been perturbatively constructed, it can be used 
in a straightforward manner for lattice simulations. 
Analytically calculated lattice-continuum relations
allow an accurate extrapolation to the continuum limit.
Thus the IR-problem plaguing direct 4d perturbative
computations can be circumvented. 

As a result of this program, the thermodynamics of the EW
phase transition is now known with a relatively good accuracy 
for a large class of theories. For the Standard Model, the
effective theory is 3d SU(2)+Higgs (neglecting the U(1) group
which has small effects), and the 
transition is too weak for baryogenesis for any $m_H$.
However, it is of the first order up to $m_H\sim 70-80$ GeV.

For the MSSM, the effective theory may be 3d SU(2)+Higgs
or more complicated, depending, in particular, on 
whether $m_{\widetilde{t_R}}\gsim 180$ GeV or not. For
the case $m_{\widetilde{t_R}}\gsim 180$ GeV
leading to eq.~\nr{L3d}, 
the available lattice results apply and baryogenesis is possible 
up to $m_H\lsim m_W$. For smaller $m_{\widetilde{t_R}}$, 
the perturbative indications are that the transition 
could be strong enough up to $m_H\lsim 100$ GeV and 
that there may even be the possibility of a two-stage transition.
However, one would need to make new lattice simulations 
to confirm this. That the transition
is stronger usually means that perturbation theory is more
reliable. However, one has seen a relatively large gauge and 
renormalization scale dependence in this regime, possibly
implying that there are surprises.

\section*{Acknowledgments}

Most of the work presented in this talk was done
in collaboration with D. B\"odeker, P. John,
K. Kajantie, K. Rummukainen, M.G.  Schmidt and/or
M. Shaposhnikov. I thank K. Kajantie for comments
on the manuscript.

\end{document}